\documentclass[10pt]{iopart}

\usepackage{graphicx}
\usepackage{setspace}
\usepackage{gensymb}
\usepackage{braket}
\usepackage{color}

\begin{document}

\newcommand{\GHzpVpcm}[0]{\mathrm{GHz\;(V/cm)^{-2}}}
\newcommand{\mVpcm}[0]{mV\;cm^{-1}}
\newcommand{\Vpcm}[0]{V\;cm^{-1}}
\newcommand{\numunit}[2]{$#1 \;\mathrm{#2}$}
\newcommand{\numunitexp}[3]{$#1 \times 10^{#2} \;\mathrm{#3}$}
\newcommand{\numunitunc}[3]{$#1 \pm #2 \;\mathrm{#3}$}
\newcommand{\state}[2]{#1_{#2}}
\def\bshift{{\sf B}}

\newcommand{\Mark}[1]{\textcolor{blue}{$\spadesuit$~#1}}

\title{Quantum computing with atomic qubits and Rydberg interactions: Progress and challenges}
\author{M. Saffman}
\address{Department of Physics, University of Wisconsin-Madison, 1150 University Avenue, Madison, Wisconsin, 53706, USA}

\begin{abstract}
We present a review of quantum computation with neutral atom qubits. After an overview of architectural options and approaches to preparing large qubit arrays we examine Rydberg mediated gate protocols and fidelity for two- and multi-qubit interactions. Quantum simulation and Rydberg dressing are alternatives to circuit based quantum computing  for exploring many body quantum dynamics. We review the properties of the dressing interaction and provide a quantitative figure of merit for the complexity of the coherent dynamics that can be accessed with dressing. We conclude with a summary of the current status and an outlook for future progress. 
\end{abstract}
\submitto{J. Phys. B}

\maketitle

\ioptwocol

\section{Introduction}

Quantum computing is attracting great interest due to its potential for solving classically intractable problems. Several physical platforms are under development and have been demonstrated at small scale  including superconductors, semiconductors, atoms, and photons\cite{Ladd2010}. Experiments with trapped ions\cite{Benhelm2008b,Harty2014,Ballance2015b} and superconducting qubits\cite{Chow2012,Barends2014} have achieved high fidelity quantum logic operations that are close to, and in some cases exceed,  known thresholds for error correcting quantum codes\cite{Devitt2013,Terhal2015}. In addition to the need for high fidelity logic gates there are several other 
requirements for translating demonstrations of quantum bits and quantum logic operations into a useful quantum computing device. These  were enumerated by DiVincenzo some years ago\cite{Divincenzo2000} and still serve as a useful check list when considering physical approaches to quantum computation. 

In this review we take a critical look at the prospects for developing scalable quantum computation based on neutral atom qubits with Rydberg state mediated entanglement. 
Although there  has been significant progress in the last year\cite{Maller2015,Jau2016},  a high fidelity two-qubit entangling gate remains to be demonstrated with neutral atoms. It is therefore tempting to focus on gate fidelity as the most important challenge facing neutral atom quantum computation. Nevertheless we will argue that gate fidelity is but one of several challenges, most of which have received much less attention than logic gate fidelity. 

The review will  be divided into sections corresponding to the DiVincenzo criteria as follows. In Sec. \ref{sec.architecture} we briefly recall the main elements of a neutral atom quantum computing
architecture. Approaches to large, scalable qubit arrays are presented in Sec. \ref{sec.array}. The important issue of trap lifetime is discussed in Sec. \ref{sec.lifetime}. In Sec. \ref{sec.coherence} we review what has been achieved for neutral atom coherence times. In Sec. \ref{sec.initializationandmeasurement} 
 we present approaches to qubit initialization and measurement with a focus on implementing these operations with low crosstalk in a multi-qubit setting.  

Section \ref{sec.gates} presents the current state of the art for neutral atom logic gates. The discussion is divided into consideration of one-qubit operations in Sec. \ref{sec.gates1} and two-qubit operations in Sec. \ref{sec.gates2}. The fundamental limits to gate fidelity  are examined in Sec. \ref{sec.intrinsic}. A particular feature of Rydberg mediated gates is the potential for multi-atom gate operations that are more efficient than a decomposition into universal one- and two-qubit gates. Section \ref{sec.gatesm} presents the protocols that have been proposed for multi-qubit operations. 
Experimental challenges for high fidelity gates are explored in Sec. \ref{sec.experimental}. 
 We conclude with a review of alternative approaches including quantum simulation, Rydberg dressing, and hybrid interactions in  Sec. \ref{sec.other} followed an outlook for the future in Sec. \ref{sec.outlook}. Primary attention is paid to developments in the last five years. A detailed presentation  of the basic ideas and earlier results on the use of Rydberg atoms for quantum information can be found in \cite{Saffman2010}.

\section{Neutral atom architecture}
\label{sec.architecture}

\begin{figure}[!t]
\centering
\begin{minipage}[c]{8.cm}
\centering
 \includegraphics[width=8.cm]{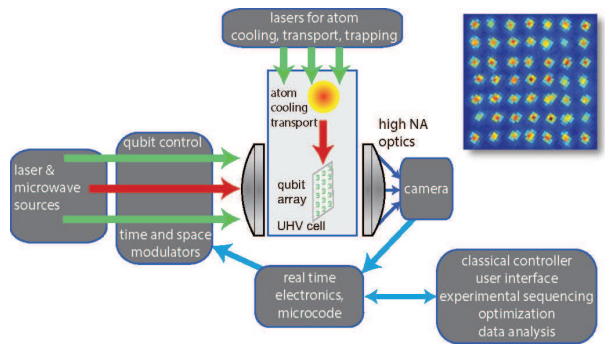}
\end{minipage}
  \caption{Architecture for a neutral atom quantum computer. The inset shows a fluorescence image of a 49 site qubit array\cite{Piotrowicz2013}. }
\label{fig.computer}
\end{figure}

Neutral atoms are being intensively developed for studies of quantum simulation\cite{Cirac2012,Bloch2012,Hauke2012} and 
computation\cite{Nielsen2000}. Aspects of quantum computation with trapped neutral atoms have been reviewed in \cite{Saffman2005a,Bloch2008,Saffman2010,Buluta2011,Negretti2011,Weitenberg2011b,Low2012,Walker2012,Lim2013,
Naber2015b,Browaeys2016}.  One vision for a neutral atom quantum computer as depicted in Fig. \ref{fig.computer} is based on an array of single atom qubits in optical or magnetic traps. 
The array is loaded from a reservoir of laser cooled atoms at $\mu\rm K$ temperature and a fiducial logical state encoded in hyperfine-Zeeman ground substates  is prepared with optical pumping. 
Logic gates are performed with some combination of optical and microwave  fields and the results are measured with resonance fluorescence. In this way all of the DiVincenzo criteria for computation can in principle be fulfilled and experiments over the last decade have demonstrated all of the required capabilities, albeit not in a single  platform, and not yet with sufficient fidelity for error correction and scalability. 

\begin{figure}[!t]
\centering
\begin{minipage}[c]{8.5cm}
\centering
 \includegraphics[width=8.5cm]{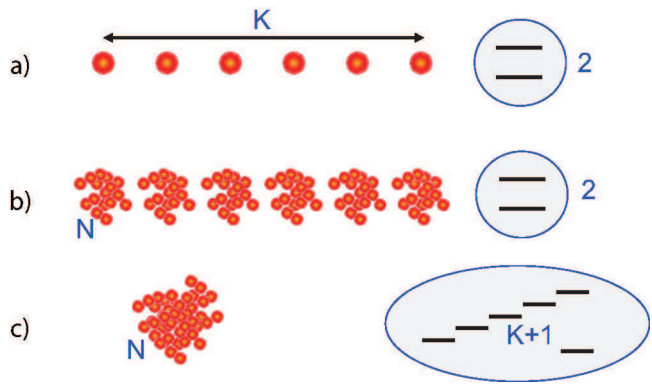}
\end{minipage}
  \caption{Encoding a $K$ qubit quantum register with neutral atoms. a) Standard method with one two-level atom per qubit. b) ensemble encoding with $N$ two-level atoms per qubit. c) Collective encoding with $K$ qubits in one ensemble using atoms with $K+1$ internal levels.  }
\label{fig.encoding}
\end{figure}

Most experimental work to date on neutral atom quantum logic has used the heavy alkalis Rb and Cs which are readily laser cooled and optically or magnetically trapped. Qubits can be encoded in Zeeman or hyperfine ground states which afford long coherence times and GHz scale qubit frequencies in the case of hyperfine qubits.    The heavy alkalis also have well resolved excited state hyperfine splittings which is important for state initialization by optical pumping and qubit measurements by resonance fluorescence. 

Quantum gates are usefully divided into one- and two-qubit operations. 
One-qubit gates on hyperfine qubits can be performed with microwaves, two-frequency stimulated optical Raman transitions, or a combination of Stark shifting light and microwaves. We defer a discussion of the current state of the art of one-qubit gate operations to Sec. \ref{sec.gates1}. Two-qubit entangling gates are possible based on several different approaches. The first demonstration of entanglement of a pair of neutral atoms used atom-photon-atom coupling between  long lived circular Rydberg states\cite{Hagley1997}. This was followed by 
lattice experiments that created entanglement between many pairs of trapped atoms using collisional interactions\cite{Mandel2003,Anderlini2007}. A recent experiment  demonstrated collisional entanglement of a single pair of atoms trapped in movable optical tweezers\cite{Kaufman2015}. In this review we will concentrate on Rydberg mediated gates\cite{Jaksch2000} which have been demonstrated in several experiments\cite{Isenhower2010,Zhang2010,Wilk2010,Maller2015,Jau2016}. The physics of the Rydberg interaction between individual atoms has been treated in detail elsewhere\cite{Gallagher1994,Choi2007,Gallagher2008,Walker2008,Comparat2010,Marcassa2014}, including a  review in this special 
issue\cite{Browaeys2016}.  Here we focus on the application of Rydberg interactions to quantum computation including  a   detailed discussion of the status and prospects for high fidelity Rydberg gates  in Sec. \ref{sec.gates}.

Several different approaches to qubit encoding are possible. Figure \ref{fig.encoding}a) shows the standard approach of encoding  a $K$ qubit register in an array of $K$ identical two-level atoms, each encoding one qubit. The Rydberg blockade interaction can be used to restrict a multi-particle ensemble to a two-dimensional logical subspace\cite{Lukin2001}. This ensemble encoding is shown in Fig. \ref{fig.encoding}b) and requires $K$ ensembles, each containing $N$ two-level atoms to encode the array. The qubit basis states in the ensemble encoding are themselves multi-particle entangled states in the physical basis. Preparation and verification of entanglement in ensemble qubits using Rydberg blockade was demonstrated recently\cite{Ebert2015,Zeiher2015}. The ensemble approach can be further extended to one $N$ atom ensemble collectively encoding $K$ qubits if each atom has at least $K+1$ internal levels and $N\ge K$\cite{Brion2007d}. Collective encoding has not yet been demonstrated experimentally but could in principle form the basis of a 1000 qubit scale experiment by coupling multiple collective ensembles\cite{Saffman2008}.

 There are several intrinsic features of neutral atoms that make them well suited for multi-qubit experiments. 
As with trapped ion qubits, neutral atoms are all identical so that the qubit frequency $\omega_q$ is the same for each and every qubit. Although the situation is more complicated when the qubits are trapped with electromagnetic fields, to first order the qubits are all identical. This is an important feature of natural, as opposed to synthetic qubits, which greatly reduces the control system complexity that is otherwise needed to address heterogeneous qubits. Not surprisingly there is also a flip side to this argument in that the identical character of atomic qubits renders them susceptible to unwanted crosstalk during preparation, logic, and measurement operations. Furthermore, in some engineered systems such as superconducting qubits, the availability of different qubit frequencies is an important feature for exercising control with low cross talk\cite{Theis2016}.

It remains a matter for debate as to whether identical or heterogeneous qubits are better suited for engineering large scale systems. 
 It has been argued recently in the context of trapped ion architectures, that identical qubits present an  advantage due to the simplified control requirements as well as better possibilities for dynamically reconfigurable qubit interconnections\cite{Brown2016}. Much the same arguments apply to neutral atom architectures, and in this section, as well as Sec. \ref{sec.initializationandmeasurement}, we highlight opportunities and challenges that exist in a neutral atom architecture based on  identical qubits.

\subsection{Qubit arrays}
\label{sec.array}

\begin{figure}[!t]
\centering
\begin{minipage}[c]{8.cm}
\centering
 \includegraphics[width=8.cm]{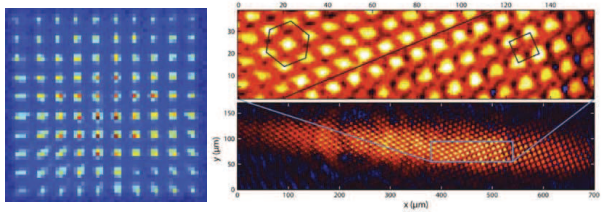}
\end{minipage}
  \caption{Fluorescence images of atoms in a 100 site 2D optical trap array (left from \cite{Nogrette2014}) and a 2D magnetic trap array (right from \cite{Leung2014}). }
\label{fig.array}
\end{figure}

Neutral atom qubit arrays may be based on trapping in optical\cite{Jessen1996,Raithel2006} or magnetic\cite{Ghanbari2006,Fortagh2007,Whitlock2009,Leung2014} lattices, examples of which are shown in Fig. \ref{fig.array}.  Due to the very weak magnetic dipole and van der Waals interactions of ground state atoms 
arrays with lattice constants of a few $\mu\rm m$ are suitable for long coherence time qubit memory while allowing site specific control with focused optical beams\cite{Knoernschild2010}, or with a gradient magnetic field\cite{Schrader2004}. In the last few years several experiments have demonstrated the ability to coherently control single atoms in 2D\cite{Weitenberg2011,Labuhn2014,Xia2015,Maller2015} and 3D\cite{YWang2015,YWang2016} arrays of optical traps. 

The number of qubits that can be implemented in a 2D or 3D array is  limited by several factors. For optical traps large arrays require more laser power. The power needed per trap site depends on the desired trap depth and the detuning from the nearest optical transitions. Small detuning gives deeper traps, with a depth scaling as $1/\Delta$, where $\Delta$ is the detuning from the nearest strong electronic transition. This must be balanced against the photon scattering rate which causes heating and decoherence and scales as $1/\Delta^2$. Arrays of up to several hundred  sites with qubit spacings of a few $\mu\rm m$ have been implemented\cite{Nelson2007,Schlosser2012,Piotrowicz2013,Nogrette2014,YWang2015,Tamura2016} and it is realistic, assuming appropriate laser development, to imagine scaling this number to $N\sim 10^4$ which  would likely be sufficient to solve 
problems in quantum chemistry that are intractable on classical machines\cite{Poulin2015}. Magnetic trap arrays have essentially no power source or dissipation limitations with respect to number of qubits, particularly if permanent magnets or superconducting wires are used. 

Preparation of single atom occupation is not deterministic when loading from a sample of laser cooled atoms. 
Single atom loading probabilities in very small trap sites are expected to be 50\% due to collisional blockade\cite{Schlosser2001}, with higher loading  possible using blue detuned catalysis light to enhance the rate of repulsive molecular interactions that only eject a single atom per collision from the trap. 
In this way 91\% loading has been achieved at a single site\cite{Carpentier2013} and 90\% in a 4 site array\cite{Lester2015}. In large arrays with blue detuned traps 60\% average
loading has been demonstrated without adding additional catalysis light\cite{Xia2015}. Loading rates above 90\% can be achieved in large arrays via the superfluid-Mott insulator transition at the expense of longer experimental cycle times\cite{Weitenberg2011}. Another approach is to start with a partially filled lattice and then rearrange the atoms to create a smaller, but fully loaded array\cite{Weiss2004}. Atom rearrangement in a 2D array to create a fully loaded 6 site subarray was recently demonstrated\cite{WLee2016} using a reconfigurable spatial light modulator to define the trapping potentials. Fast loading  of  arrays  with $\sim 50 $ single atoms to filling fractions close to 100\% on $\sim 100~\rm  ms$ timescales has been recently achieved in 1D\cite{Endres2016} and 2D\cite{Barredo2016} geometries using movable optical tweezers.

Irrespective of the number of traps that can be implemented and filled any technology will also be limited by the number of traps that can be addressed and controlled. Optics provides the wiring for atomic qubits which can be an advantage compared to electronics since optical beams can be rapidly reconfigured and qubits do not need to be tethered to physical wires which can introduce decoherence. At the same time optical technologies are less developed than electronic circuits and components, and the technology base needed for optical qubit control is not yet sufficiently advanced. Preparation of patterned arrays in short period lattices has been demonstrated with optical\cite{Weitenberg2011} and electron beam\cite{Wurtz2009} addressing.  The most promising approaches for coherent single site addressing in large arrays rely on 
scanning of focused optical beams using either electro-optic modulators\cite{Schmidt-Kaler2003}, acousto-optic modulators\cite{Nagerl1999,Kim2008extra,Xia2015}, piezo mirrors\cite{Weitenberg2011b}, micro-electro-mechanical devices\cite{Knoernschild2010,YWang2015}, or spatial light  modulators\cite{vanBijnen2015}. With all of these technologies there are trade offs in the design space between speed, crosstalk, and number of addressable sites. Acousto-optic devices can have space-bandwidth products of several thousand and crossed devices for 2D addressing\cite{Xia2015,Maller2015} have the potential to control $N=10^4$ sites. 
Also digital mirror device (DMD) spatial light modulators such 
as were used in \cite{Pasienski2008} have great potential for addressing large qubit arrays. Commercially available devices have more than $10^6$ pixels, and full frame switching speeds of $\sim 30~\mu\rm s$. While this speed is not competitive with electro-optic or acousto-optic modulators one can envision architectures that leverage the large pixel count, combined with several modulators, to achieve fast and simultaneous addressing of multiple sites.

\subsection{Trap lifetime}
\label{sec.lifetime}

Any scalable quantum computing technology will undoubtedly rely on quantum error correction techniques.  
When considering a neutral atom approach it is necessary to recognize that atom loss represents an unavoidable component of the neutral atom error budget. 
 The longest neutral atom trap lifetimes reported to date are about 1 hour in a cryogenic environment\cite{Willems1995}. The trap lifetime is limited by collisions with untrapped background atoms. To suppress the loss rate due to collisions it is necessary to raise the trap depth to be comparable to, or even larger than,  $k_B T_{\rm background}$ the energy of untrapped atoms. While there is no fundamental limit to the lifetime with sufficiently deep traps  there are practical limitations.

Optical traps for neutral atoms cannot be arbitrarily deep since both the trap depth and the photon scattering rate, which causes decoherence, scale proportional to the optical power. Estimates with nominal atomic parameters show that traps as deep as a few times 10 mK are feasible. For example a Cs atom trapped by a $\lambda=1.06~\mu\rm m$ laser with an intensity sufficient for a 20 mK trap depth implies a Rayleigh scattering rate of about $220~\rm s^{-1}$ and a Raman scattering rate of about  $3~\rm s^{-1}$. The power required in a $ 1 ~\mu\rm m$ beam waist is about 140 mW.
Scaling the trap depth to 300 K is impractical and even scaling to 4 K would require more than 1 W of power for each trap site. 
If an array of such traps were used in a dilution refrigerator with $T_{\rm background}\sim 10 ~\rm mK$ extremely long collision limited lifetimes should be possible. However, for more than a few trap sites one would need Watt scale power levels while available large dilution refrigerators have cooling power not more than 1-2 mW. Thus there would be extreme technical challenges associated with residual  optical scattering and absorption leading to unmanageable heat loads. While other trap wavelengths as well as blue detuned traps can be considered, creating an array of 10 mK deep optical traps inside a dilution refrigerator while providing all the optical access needed for qubit control appears to be a very difficult challenge. 

Magnetic traps provide an interesting alternative.  For alkali atoms in the most magnetically sensitive stretched ground state a 4 K trap depth requires a peak field of about 6 T. It is difficult to envision modulating such large fields on the few $\mu\rm m$ length scales desirable for qubit arrays. On the other hand a  10 mK magnetic trap would require  only about 15 mT of peak field strength.  A $\mu\rm K$ temperature atom  localized near the field minimum in a  Ioffe-Pritchard type trap would be subjected to even smaller fields.  While substantial challenges remain as regards optical access and minimization of light scattering inside a dilution refrigerator we conclude that magnetic trap arrays in a cryogenic environment 
could in principle provide a scalable platform with very long qubit lifetimes.

It should also be mentioned that even in the absence of collisional losses atom heating due to fluctuations of the trapping potential can eventually cause trapped atoms to escape. Heating mechanisms in optical traps have been discussed in \cite{Gehm1998}. In principle trap heating does not impose a fundamental limit since it is possible to periodically recool trapped atoms, thereby increasing the lifetime\cite{Gibbons2008}.

\begin{figure}[!t]
\centering
\begin{minipage}[c]{8.cm}
\centering
 \includegraphics[width=8.cm]{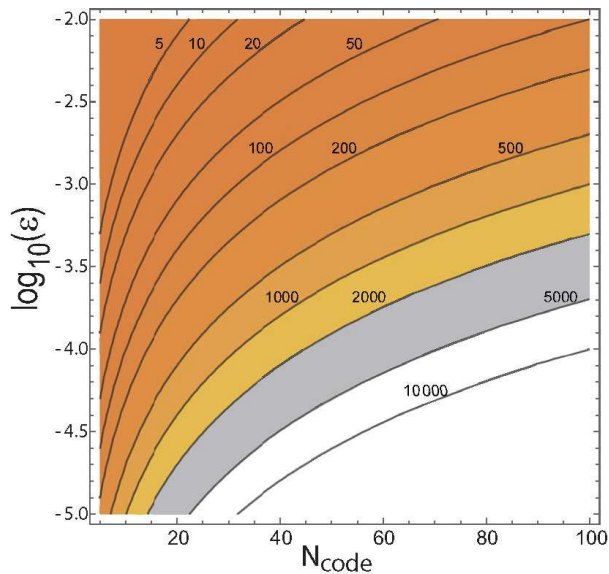}
\end{minipage}
  \caption{Contours labeled with $\tau_{\rm vac}$ in seconds from Eq. (\ref{eq.lifetime}) as a function of code size $N_{\rm code}$ and error 
threshold $\epsilon$ assuming $t_{\rm qec} = 0.1 N_{\rm code} ~\rm (ms)$. }
\label{fig.lifetime}
\end{figure}

Let's estimate relevant error numbers due to qubit loss. 
Consider a quantum error correcting code that  employs $N_{\rm code}$ single atom qubits and the vacuum limited lifetime of each qubit is $\tau_{\rm vac}$. Then the probability of having lost 
at least one bit after time $t$ is $P_{\rm loss}=N_{\rm code} (1-e^{-t/\tau_{\rm vac}})\simeq N_{\rm code} t/\tau_{\rm vac}.$
In order for the probability to be less than $\epsilon$ for a quantum error correction cycle of duration $t_{\rm qec}$ we require 
\begin{equation}
\tau_{\rm vac} > \frac{N_{\rm code} t_{\rm qec}}{\epsilon}.
\label{eq.lifetime}
\end{equation}
If an atom loss error is to be correctable then (\ref{eq.lifetime}) must be satisfied with  $t_{\rm qec}$ corresponding to the time needed to both diagnose and correct an atom loss event. Atom loss can be detected using a ``knock-knock" protocol introduced by Preskill\cite{Preskill1998}. A missing atom can then be replaced using transport from a reservoir of cold atoms. Single atom transport has been demonstrated in several 
experiments\cite{Nussmann2005,
Fortier2007,Khudaverdyan2008,Thompson2013,Dinardo2015} as well as the recent assembly of arrays with high filling 
factor\cite{Endres2016,Barredo2016}. Figure \ref{fig.lifetime} shows the necessary values of $\tau_{\rm vac}$ 
assuming  $t_{\rm qec} = 0.1 N_{\rm code} ~\rm (ms)$, so a 10 qubit code can be checked and repaired in 1 ms. 

We see that a moderate sized code word of 20 qubits, which counts both data and ancillas, and a threshold of $\epsilon=0.0001$ would require $\tau_{\rm vac}=400~\rm s$ which is certainly achievable in a well designed ultra high vacuum system. To estimate the required rate of atom reloading consider a 100 logical qubit  processor with $N_{\rm code}=20$ and  $N_{\rm phys}=2000$ physical qubits. Using $t_{\rm qec} \ge 1/r_{\rm load}$, where $r_{\rm load}$ is the time required to reload a lost atom, the condition (\ref{eq.lifetime}) can be expressed as
 \begin{equation}
r_{\rm load} > \frac{N_{\rm phys} }{\tau_{\rm vac} \epsilon}.
\label{eq.lifetime2}
\end{equation}
Using $N_{\rm phys}=2000$, $\tau_{\rm vac}=400~\rm s$, and $\epsilon=0.0001$ we find $r_{\rm load}>5\times 10^4 ~\rm s^{-1}$. Array assembly experiments\cite{Endres2016,Barredo2016} have demonstrated $r_{\rm load}\sim 10^3~\rm s^{-1}$ in 1D (2D).  With further improvements in loading rate, which may require additional parallelization of multiple atom transport beams,  continuous operation and error correction of an array with $N_{\rm phys}\sim 2000$ will become feasible.

The simple estimates of Eqs. (\ref{eq.lifetime},\ref{eq.lifetime2}) show that small code sizes, fast measurements and atom replacement, together with long lifetimes will be essential for continuously operating neutral atom quantum logic. These requirements suggest that future implementations may advantageously be placed in a cryogenic environment to increase atom lifetimes. This has the secondary advantage when considering Rydberg gates that Rydberg lifetimes will also be lengthened which will increase gate fidelity (see Fig. \ref{fig.lifetime}). 
In addition to the above requirements it will be necessary to implement a control system that can diagnose and correct errors on multiple code blocks in parallel. Sequential monitoring of a large array with say $N=10^4$
qubits would imply impractically long atom lifetimes.

An alternative approach to correcting for atom loss uses an ensemble encoding of a qubit in multiple atoms\cite{Lukin2001}. 
Since loss of one atom only causes ${\mathcal O}(1/N)$ fidelity loss in an $N$ atom ensemble qubit, it is possible to correct the encoded states against loss\cite{Brion2008}. 
 The central challenge of an ensemble approach is reaching high enough gate fidelity to make logical encoding feasible. Experimental results to date indicate worse gate fidelity and shorter coherence times  for ensemble qubits than for single atom 
qubits\cite{Ebert2014,Ebert2015,Zeiher2015,Weber2015}.  
This is not surprising due to the sensitivity of a delocalized ensemble to field gradients as well as the presence of  atomic collisions and possibly molecular resonances\cite{Derevianko2015}. These effects may be mitigated by patterned loading of a localized ensemble into an optical lattice to prevent short range interactions as described in \cite{Saffman2008}.

\section{Coherence}
\label{sec.coherence}

Long coherence times have been demonstrated with atomic hyperfine qubits. For the purpose of comparison it is useful to consider longitudinal relaxation ($T_1$) and transverse relaxation ($T_2$) times separately.  

The $T_1$ time for trapped atoms is sensitive to external fields and light scattering. Collisional loss rates, as have been discussed in the preceding section, set an upper limit on  the effective $T_1$. Nevertheless the  $T_1$ limit  due to external fields tends to be much shorter than the collisional loss times in an ultrahigh vacuum environment. For qubits encoded in Zeeman substates with MHz scale energy separations magnetic field noise can cause transitions between states. For the more common case of encoding in different hyperfine levels the qubit energy separation is several GHz in alkali atoms and transitions caused by low frequency magnetic fields are negligible. Unshielded microwave frequency fields can cause transitions, and rf shielding around the experiment is important. 

The transverse coherence time $T_2$ is sensitive to dephasing mechanisms that preserve population but randomly change the relative phase of the qubit  states. Dephasing occurs due to magnetic noise, intensity noise in optical traps, and due to motional effects for atoms that are not cooled to the vibrational ground state of the trapping potential\cite{Kuhr2005}. Due to the importance of coherence for qubit experiments, and for atomic clocks, this problem has received a great deal of attention\cite{Wineland1998,Ye2008}. With appropriate choices for the  hyperfine Zeeman states, and settings  of the optical intensity, polarization, and  magnetic field it is possible to store atomic states with long coherence times of many seconds\cite{Treutlein2004,Lundblad2010,Chicireanu2011,Dudin2010,Dudin2013,Derevianko2010a,Carr2014,JYang2016}.

\subsection{Rydberg magic trapping}

Rydberg gates require excitation to atomic levels that may have different trapping potentials than the qubit ground states. Heating, or anti-trapping, due to the change of potential can be mitigated by turning the trap off for the short duration of the Rydberg interaction. Although this has  been the standard approach for experiments with a few atoms it will be advantageous in future array experiments to be able to perform gates while keeping the trap on. To minimize heating and decoherence rates the traps should be designed to satisfy a ground-Rydberg magic condition.  Early work towards designing such traps\cite{Safronova2003,Saffman2005a} was reviewed in \cite{Saffman2010}, Sec. IV.A.2. The approach of \cite{Saffman2005a} was implemented in an experiment that demonstrated entanglement between light and Rydberg excited atoms\cite{LLi2013}.

\begin{figure}[!t]
\centering
\begin{minipage}[c]{8.cm}
\centering
 \includegraphics[width=8.cm]{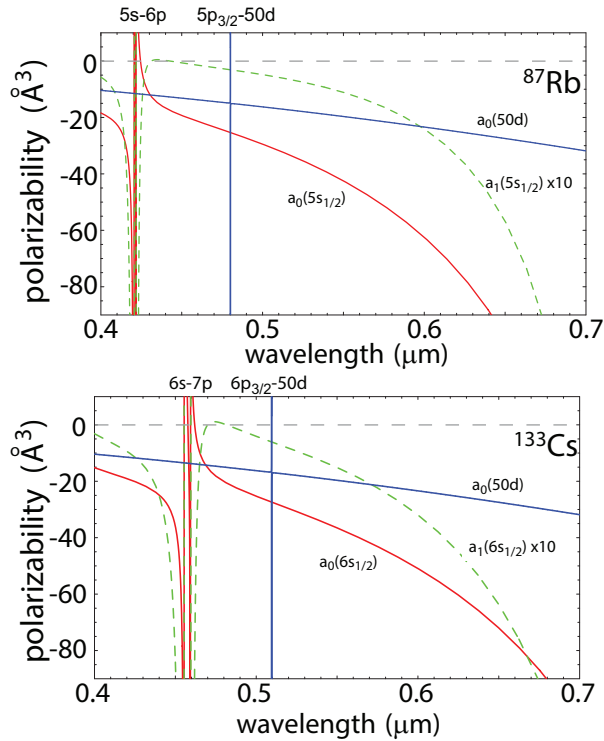}
\end{minipage}
  \caption{Scalar polarizability of ground states and
Rydberg states of Rb (top) and Cs (bottom) from \cite{SZhang2011}. The ground state vector  polarizability is shown as a dashed line.
}
\label{fig.polarizability}
\end{figure}

The early proposals for ground-Rydberg magic traps mentioned above had the drawback that they required relatively small detuning from transitions originating in either the ground or Rydberg states. This leads to excessive scattering rates and suboptimal coherence properties. Later work has shown that magic traps can be designed to work over a broad wavelength range by correct matching of the shape and size of the trapping potential to the Rydberg electron wavefunction.  
To a first approximation  the wavefunction of the Rydberg excited valence electron is close to that of a free electron and therefore has a negative polarizability $\alpha_{\rm Ryd}=-e^2/(m\omega^2<0$ where $e,m$ are the electron charge and mass and $\omega$ is the frequency of the light.
This free electron ponderomotive potential was suggested for optical trapping of Rydberg atoms in\cite{Dutta2000}.  Refined calculations have verified the accuracy of the free electron polarizability at wavelengths away from transition resonances to better than 1\% in high $n$ Rydberg atoms\cite{Topcu2013b}. 
Choosing a wavelength such that also the ground state atom has a negative polarizability leads to the same sign of the trapping potential. Atoms with negative polarizability can be confined in dark traps that have an intensity minimum surrounded by light. Since the Rydberg wavefunction is delocalized the Rydberg electron will see a larger intensity than the ground state atom and potential balancing requires that $|\alpha_{\rm Ryd}|<|\alpha_{\rm ground}|$ with $\alpha_{\rm Ryd},\alpha_{\rm ground}<0.$ These conditions are satisfied over a broad wavelength range, as can be seen in Fig. \ref{fig.polarizability}. Detailed analysis in \cite{SZhang2011} verified the possibility of magic trapping for a wide range of wavelengths. Measurements of the state dependent ponderomotive potential for Rydberg atoms have been reported in \cite{Younge2010} and one-dimensional Rydberg trapping in \cite{Anderson2011}.

Somewhat less obvious is the possibility of magic trapping at intensity maxima with wavelengths for which the ground state has a positive polarizability, $\alpha_{\rm ground}>0.$ This is possible even though $  \alpha_{\rm Ryd}<0$ by taking into account the large spatial extent of the Rydberg wavefunction. Careful consideration of the three dimensional overlap of the  wavefunction with the repulsive ponderomotive potential leads to a ``landscape modulated" trap for which there is high trap intensity inside the electron distribution\cite{Topcu2013}. Unfortunately this only occurs for long trap wavelengths near $10~\mu\rm m$ and high principal quantum number (for example $n\ge 154$  for Rb $ns$ states) so the applicability to array experiments is uncertain. Additional studies have pointed out the possibility of simultaneous magic trapping of both qubit states and a Rydberg state using Al\cite{Morrison2012} or divalent atoms\cite{Topcu2014,Topcu2016}.

Finally we mention that magic magnetic trapping of ground and Rydberg states is also a possibility. For example a qubit could be encoded in alkali atom states $f=I+1/2,m_f=1$ and $f=I-1/2, m_f=-1$ ($I$ is the nuclear spin) which are both low field seeking. Low angular momentum Rydberg states can be trapped in Ioffe-Pritchard potentials if the  Zeeman sublevel is chosen to be a low field seeker\cite{Mayle2009}.
In principle it may be possible to match the trap potentials, although detailed calculations have not been performed.

\section{Initialization and Measurement}
\label{sec.initializationandmeasurement}

Scalable quantum computation relies on a combination of coherent and dissipative processes. While gate operations typically rely on coherent dynamics, qubit initialization and measurements, which are required for error correction, are  dissipative processes that remove entropy by coupling qubits to radiation fields. Initialization can be performed by optical pumping, and state measurements by detection of resonant fluorescence, although absorptive or phase shift measurements with focused probe beams are also possible\cite{Wineland1987,Aljunid2009}. 

One of the outstanding challenges  is implementation of quantum nondemolition (QND) qubit state measurements without loss. Strictly speaking quantum computation and error correction could proceed with a QND measurement capability. However, the overhead required to correct for lossy measurements is prohibitive.  The most widely used approach for  qubit measurements with neutral alkali atoms relies on imaging of fluorescence photons scattered from a cycling transition between one of the qubit states and the strong D2 resonance 
line\cite{Saffman2010}. Due to a nonzero rate for spontaneous Raman transitions from the upper hyperfine manifold there is a limit to how many photons can be scattered, and detected, without changing the quantum state. This problem is typically solved by preceding a measurement with resonant ``blow away" light that removes atoms in one of the hyperfine states. The presence or absence of an atom is then measured with repumping light turned on, and a positive measurement result is used to infer that the atom was in the state that was not blown away.

This method can indeed provide high fidelity state measurements but has several drawbacks. An atom is lost half the time on average, and must be reloaded and reinitialized for a computation to proceed. Atom reloading involves mechanical transport, and thus tends to be slow compared to gate and measurement operations. 
 Lossless QND measurements that leave the atom in one of the qubit states, or at least in a known Zeeman sublevel of the desired hyperfine state, can be performed provided that the measurement is completed while scattering so few photons that the probability of a Raman transition is negligible. This was first done for atoms strongly coupled to a cavity\cite{Boozer2006,Bochmann2010,Gehr2010}, and was subsequently   extended to atoms in free space\cite{Gibbons2011,Fuhrmanek2011,Jau2016} using photon counting detectors. 

Multiplexed QND state detection of more than two atoms in an array using an imaging detector has not yet been demonstrated. Array measurements are typically performed with electron multiplying charge coupled device (EMCCD) cameras which suffer from excess noise compared to discrete photon detectors\cite{Alberti2016}. Recent progress has resulted in initial demonstrations of EMCCD based QND state measurements\cite{Martinez-Dorantes2016,Kwon2016} which establishes a path towards fast measurement of large qubit arrays.

Despite progress towards QND state measurements there is an outstanding challenge related to crosstalk during measurements. 
It appears necessary for error correction that at least one of the operations of initialization or measurement can be performed with low crosstalk to other qubits in a multi-qubit experiment. While both of these operations can in principle be performed at selected trap sites using tightly focused beams crosstalk lurks due to the large resonant cross section of proximal atoms. 
To put the situation in perspective we can make some estimates as follows. The resonant cross section for photon absorption is $\sigma=\frac{3}{2\pi}\lambda^2$ so with qubits in recent lattice experiments  spaced by $d\sim 5\lambda$\cite{Xia2015,YWang2015} the probability of a scattered photon being absorbed is $\eta_{\rm abs}\sim \sigma/(4\pi d^2)\sim 0.0015.$ If the qubit measurement is performed with a moderately high numerical aperture collection lens of $NA=0.5$ and the optical and detector efficiencies are 50\% the probability of photon detection is $\eta_{\rm det}\sim 0.034$ so that $\eta_{\rm abs}/\eta_{\rm det}\sim  0.04$. This ratio implies that  a   state measurement based on detection of only a single photon would incur a $\sim 4\%$ probability of unwanted photon absorption at a neighboring qubit. This 4\% error rate is too large to be efficiently handled by protocols for quantum error correction. While this estimate can be improved on by increasing $d$ or increasing the NA of the detection lens it appears difficult to reduce  crosstalk errors below thresholds for error correction. 

There are several possible routes to mitigating crosstalk in neutral atom array experiments. 
One idea is to use a focused beam to locally Stark shift the optical transition in the qubit to be 
initialized or measured so that scattered photons are detuned from neighboring qubits. Calculations indicate that this can add a suppression factor of $>100$ relative to the case of no Stark shifting\cite{Carr2014t}. Another possibility is to shelve surrounding qubits in hyperfine states that are detuned by a qubit frequency\cite{Beterov2015}. Since alkali atoms do not possess electronically excited metastable states the shelving procedure is relatively complex.

\begin{figure}[!t]
\centering
\begin{minipage}[c]{8.cm}
\centering
 \includegraphics[width=8.cm]{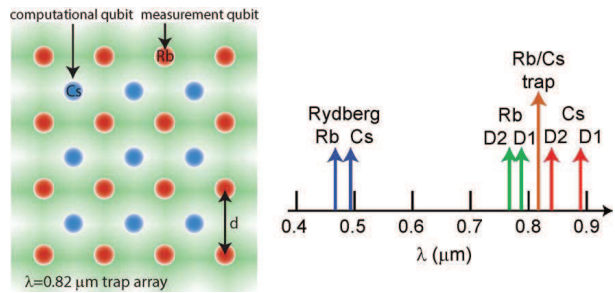}
\end{minipage}
  \caption{Proposed dual species experiment formed by a Gaussian beam array\cite{Piotrowicz2013} with wavelength $\lambda=0.82~\mu\rm m$ which traps Cs atoms in dark traps and Rb atoms in bright traps on interleaved square lattices (from \cite{Beterov2015}). The wavelengths needed for cooling, initialization, measurement, and two-photon Rydberg excitation via the D2 transitions are shown on the right.}
\label{fig.dualspecies}
\end{figure}

Yet another way of evading crosstalk is with a two-species architecture as proposed in \cite{Beterov2015}, and shown in Fig. \ref{fig.dualspecies}.  Cs atom qubits are used for computation and Rb atoms are used as auxiliary qubits for measurement. To perform a measurement on a Cs qubit a gate is performed to transfer the Cs state to proximal Rb qubits which can then be read out via resonance fluorescence. After the measurement the Rb qubits are reset with optical pumping. This is possible due to the presence of interspecies F\"orster resonances which provide strong coupling between Cs and Rb Rydberg atoms\cite{Beterov2015}. The large wavelength separation of the D1 and D2 resonance  lines in Cs and Rb give a large suppression of crosstalk 
for such an architecture. 
 
While there are several potential solutions towards crosstalk free initialization and measurement, none of them have as yet been demonstrated experimentally. Doing so will constitute an important step towards scalable quantum computation.

\section{Quantum gates}
\label{sec.gates}

Computation relies on the availability of high fidelity quantum gate operations. These can be divided into one- and two-qubit gates. As is well known\cite{Barenco1995}  a universal set of elementary one- and two-qubit gates allows for universal computation on $N$ qubits. We discuss the current status of one-qubit gates in Sec. \ref{sec.gates1} and two-qubit gates in Sec. \ref{sec.gates2}. In addition long range Rydberg interactions can be used to efficiently implement multi-qubit gate operations. While multi-qubit gates can always be decomposed into one- and two-qubit gates, there can be large improvements in efficiency and error tolerance by using native multi-qubit operations. These are discussed in Sec. \ref{sec.gatesm}. Although gate protocols have been developed that promise high fidelity compatible with scalable architectures there are a plethora of technical challenges that remain to be solved. An overview of these issues is provided in Sec. \ref{sec.experimental}.

\subsection{One-qubit gates}
\label{sec.gates1}

Single qubit gates acting on qubits encoded in atomic hyperfine states can be performed with microwaves\cite{Schrader2004,Olmschenk2010}, with two-frequency Raman light\cite{Yavuz2006,Knoernschild2010}, or with a combination of microwaves and a gradient field for addressing of individual qubits\cite{Dotsenko2004,Xia2015,YWang2015,YWang2016} or groups of qubits\cite{Lee2013}. The most recent experiments have provided detailed characterization of gate fidelity at Stark shift selected sites in large multi-qubit arrays. Using randomized benchmarking\cite{Knill2008}  average fidelities for Clifford gates of $0.992$\cite{Xia2015} and $0.996$\cite{YWang2016} have been demonstrated. Crosstalk errors to nearby, non-targeted qubits were $0.014$\cite{Xia2015} and $0.0046$\cite{YWang2016}. Improved gate fidelity and reduced crosstalk were demonstrated in \cite{YWang2016} by implementing a sequence of pulses which make  gate errors sensitive to the fourth power of small beam pointing errors.  

The error budgets in these experiments are associated with inhomogeneities in the microwave field, variations in trap induced qubit frequency shifts, and errors from the Stark addressing beams due to imperfect spatial addressing,
leakage to nontargeted sites, and residual light scattering. Reduced sensitivity to pointing errors together with reduced leakage to other sites can  be achieved by spatial shaping of the Stark beam\cite{Gillen-Christandl2016}.  While much work remains to be done, it should be possible to reduce  single qubit gate errors to  $\sim 10^{-4}$ and below, a level of performance that has already been demonstrated with trapped ion hyperfine qubits\cite{Harty2014,Mount2015}.

\subsection{Two-qubit gates}
\label{sec.gates2}

\begin{figure}[!t]
\centering
\begin{minipage}[c]{8.5cm}
\centering
\includegraphics[width=8.5cm]{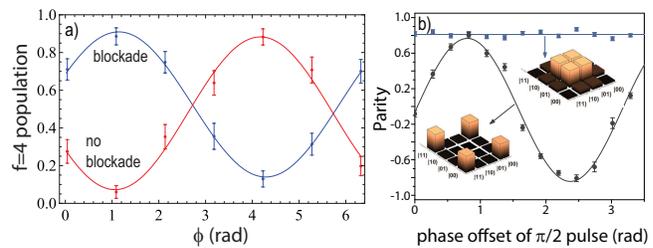}
\end{minipage}
  \caption{Rydberg gate experiments with Cs atoms. a) Eye diagram of CNOT gate with and without blockade as a function of the relative phase $\phi$ between $\pi/2$ pulses on the target qubit from \cite{Maller2015}. b) Parity oscillations of Bell states from \cite{Jau2016}. }
\label{fig.entangle}
\end{figure}

Two-qubit entanglement and logic gates using Rydberg state interactions have been demonstrated in several experiments and are listed in Table \ref{tab.gate}. 
The first Rydberg blockade entanglement experiments were performed in 2010\cite{Wilk2010,Isenhower2010,Zhang2010}. These were followed by improved results in 2015\cite{Maller2015,Jau2016}. Experimental gate results are shown in   Fig. \ref{fig.entangle}.  Two-qubit entanglement was also achieved using local spin exchange with atoms in movable tweezers in 2015\cite{Kaufman2015}, but with lower fidelity than the Rydberg experiments.   While single qubit gate operations with neutral atom qubits have already reached high fidelity, as discussed in Sec. \ref{sec.gates1}, there is a large gap between the  fidelity results summarized in Table \ref{tab.gate} and the very high entanglement fidelities that have been demonstrated with trapped ion\cite{Benhelm2008b,Harty2014,Ballance2015b} and superconducting\cite{Barends2014,Chow2012,Corcoles2015} qubits.

\begin{table*}[!t]
\caption{Experimental Bell state fidelity achieved in two-qubit neutral atom experiments. All fidelities were measured using parity oscillations\cite{Sackett2000}. Values in parentheses are less than the sufficient  entanglement threshold of 0.5, but may still represent entangled states as was explicitly verified in\cite{Kaufman2015}. Post selected values are corrected for atom loss during the experimental sequence. Experiments were performed in the year indicated. 
   }
\vspace{-.0cm}
\begin{center}
%\begin{ruledtabular}
\begin{tabular}{c|l|c|c|c}
 atom  & method & Bell fidelity & post selected fidelity & year \& reference\\ 
\hline
\hline
$^{87}$Rb&blockade, simultaneous addressing& (0.46)  & 0.75 &2009 \cite{Wilk2010}\\
$^{87}$Rb&blockade, separate addressing& (0.48) & 0.58 &2009 \cite{Isenhower2010}\\
$^{87}$Rb&blockade, separate addressing& 0.58& 0.71 &2010 \cite{Zhang2010}\\
Cs&blockade, separate addressing& 0.73 &0.79  &2015 \cite{Maller2015}\\
Cs&dressing, simultaneous addressing& 0.60 &0.81 &2015 \cite{Jau2016}\\
$^{87}$Rb&local spin exchange& (0.44) & 0.63&2015 \cite{Kaufman2015}\\
\end{tabular}
%\end{ruledtabular}
\end{center}
\label{tab.gate}
\end{table*}

Scalable computation requires quantum error correction (QEC) with the gate fidelity needed for QEC dependent on the size and architecture of the code blocks. Roughly speaking larger codes can tolerate gates with higher errors\cite{Devitt2013,Terhal2015}, with large scale surface codes that combine hundreds of physical qubits to create a single logical qubit having threshold error rates  $\sim0.01$.  The requirement of managing atom loss in a neutral atom qubit array, see Fig. \ref{fig.lifetime}, suggests that smaller code sizes are preferable. Concatenated codes with sizes of 25 qubits or less have thresholds $\sim0.001$ and for scalability gate error rates should be at least a factor of 10 better. We conclude that scalable neutral atom quantum computing will require a two-qubit gate fidelity  of   $F\sim 0.9999$. This is not a fundamental limit, and could be relaxed with the invention of efficient codes that tolerate higher errors, but serves as a placeholder with which to evaluate the status of current approaches. 

Comparing this target performance with Table \ref{tab.gate} it is apparent that in order for Rydberg gates to be viable for scalable quantum computation the fidelity  needs to be greatly improved. It is therefore important to identify the reasons for the relatively low fidelity demonstrated to date. There are two aspects of gate fidelity that can be considered separately. The first is the intrinsic gate fidelity, set by the relevant Hamiltonian, which could be achieved with a perfect experimental apparatus. The second aspect is identifying experimental imperfections that reduce the gate fidelity below the intrinsic limit.

\subsection{Intrinsic gate fidelity}
\label{sec.intrinsic}

The Rydberg blockade entangling gate introduced in \cite{Jaksch2000} relies on a three pulse sequence: control qubit $\pi$ pulse $\ket{1}\rightarrow\ket{r}$,  target qubit $2\pi$ pulse $\ket{1}\rightarrow\ket{r}\rightarrow\ket{1}$,  control qubit $\pi$ pulse $\ket{r}\rightarrow\ket{1}$,
with $\ket{1}$ the qubit level that is resonantly excited to a Rydberg state $\ket{r}$. Due to Rydberg-Rydberg interactions the state $\ket{rr}$ experiences a blockade shift   $\sf B$  relative to the singly excited states $\ket{1r}, \ket{r1}$. In the ideal limit where ${\sf B},\omega_q \gg \Omega\gg 1/\tau$ with $\omega_q$ the frequency splitting of the qubit states, and $\tau$ the radiative lifetime of the Rydberg level this pulse sequences provides an ideal entangling phase gate $C_Z={\rm diag}[1,-1,-1,-1]$. Accounting for imperfect blockade and radiative decay the  smallest gate error is achieved by setting the Rydberg excitation frequency to\cite{Saffman2005a,Saffman2010}
$\Omega_{\rm opt}=(7\pi)^{1/3}\frac{\bshift^{2/3}}{\tau^{1/3}}$ which gives an error estimate for the CNOT truth table assuming perfect single qubit operations of 
\begin{equation}
E_{\rm min} =\frac{3(7\pi)^{2/3}}{8}\frac{1}{(\bshift\tau)^{2/3}}.
\label{eq.Emin}
\end{equation}

The van der Waals interaction results in a blockade shift  $\bshift\sim n^{11}-n^{12}$  and the lifetime scales as $\tau\sim 1/(1/\tau_0 n^3 +1/\tau_{\rm BBR})$ for low angular momentum states of the heavy alkalis\cite{Beterov2009,Beterov2009b}. The $\tau_0, \tau_{\rm BBR}$ factors account for spontaneous and blackbody induced transitions.  Thus Eq. (\ref{eq.Emin})  naively suggests that the intrinsic gate error can be made arbitrarily small by using large principal quantum numbers $n$ for the Rydberg state. Unfortunately this scaling breaks down at high $n$ since the spacing of neighboring levels is $\delta U= E_{\rm H}/n^3$ with $E_{\rm H}$ the Hartree energy,  neglecting the corrections from quantum defects. Effectively the blockade is limited to the smaller of $\delta U/2$ and $\bshift$ so the minimum gate error switches at high $n$ to  $E_{\rm min}\sim 1/(\delta U \tau)^{2/3}$, which implies a slow increase of the gate error with $n$. Putting the atoms in a low temperature cryostat with 
$k_B T \ll \delta U$ increases the lifetime to $\tau\sim n^3$ in which case the gate error is asymptotically independent of $n$ with the limiting value of $E_{\rm min} =\frac{3(14\pi)^{2/3}}{8}(\hbar/E_{\rm H}\tau_0)^{2/3}$. The $np$ alkali states have the longest lifetimes, and using $\tau_0=3.3~\rm ns$ for Cs  we find $E_{\rm min}\simeq 2\times 10^{-5}$.
  
This error floor is below our scalability target of $F\sim 0.9999$ but is only relevant when the  input state is an element of the two-qubit computational basis. Inputs that are in superposition states may result in entangled output states, which suffer additional errors due to phase variations that are not accounted for by the estimate (\ref{eq.Emin}). 
A detailed determination of the gate error when creating entangled states requires following the  coherent evolution and inclusion of leakage to neighboring Rydberg levels which was first done in \cite{XZhang2012}. Numerical results based on a master equation simulation of process tomography for Rb or Cs atoms showed that in a 300 K bath the process fidelity of the $C_Z$  gate  using $ns, np,$ or $nd$ Rydberg states that can be reached by one- or two-photon transitions from the ground state is at best 0.9989 on the basis of the calculated fidelity  and 0.9988 on the basis of the trace distance using a modified gate sequence with a phase modulated Rydberg pulse to compensate for the leading order phase error from imperfect blockade\cite{XZhang2012}.  In a 4 K environment with reduced blackbody induced depopulation of the Rydberg states the gate fidelity was predicted to improve to slightly better than 0.999.  

These intrinsic limits are an order of magnitude worse than the target fidelity of 0.9999 outlined above. A large amount of work has been devoted to analyzing alternative ideas with improved fidelity.  Circular Rydberg states in a cryogenic environment have lifetimes that scale as $n^5$, substantially longer than low angular momentum Rydberg states. The analysis in \cite{Xia2013} showed that if circular states could be excited on fast time scales with low errors then gate errors $<10^{-5}$ would be theoretically possible. Unfortunately excitation of circular states requires a high order multiphoton transition so the experimental challenges are daunting.  

Other work has analyzed the use of off-resonant Rydberg excitation\cite{Brion2007b}, optimal control pulse shapes for the Rydberg excitation\cite{Muller2011,Muller2011b,Goerz2014,Muller2016}, adiabatic gate protocols\cite{Moller2008,Muller2014,Petrosyan2014,Rao2014,Tian2015,Beterov2016b},
a spin-exchange version of the blockade gate\cite{Shi2014}, an asymmetric interaction phase gate\cite{Wu2010}, a gate involving superpositions of multiple Rydberg levels\cite{XShi2016}, and a two-species interaction gate where the second species acts to mediate long-range gates beyond the range of the direct interaction of a single species\cite{Wade2016}. While optimal control and adiabatic protocols can lead to improved robustness with respect to parameter variations, as well as a reduced requirement for strong Rydberg interactions, the fidelity has been limited to less than 0.999 when accounting for Rydberg state decay. It is also possible to implement a Rydberg gate using a single continuous Rydberg pulse applied to both qubits, but without exciting the doubly occupied Rydberg state $\ket{rr}$ so there are no interatomic forces during the gate\cite{Jaksch2000,Han2016,Su2016}.  Such protocols reduce the complexity of the pulse sequence but have not been shown to reach fidelity better than 
0.997.

\begin{figure}[!t]
\centering
\begin{minipage}[c]{8.5cm}
\centering
\includegraphics[width=8.5cm]{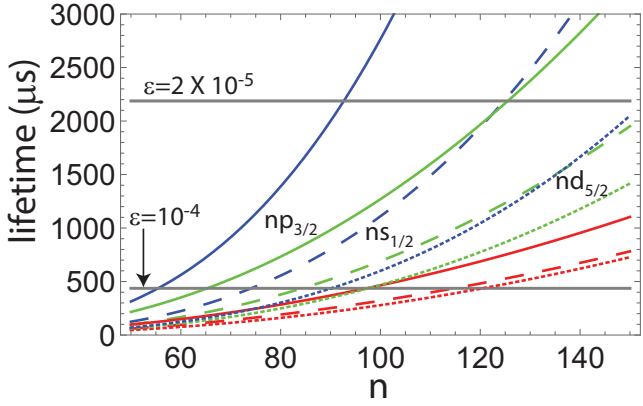}
\end{minipage}
  \caption{Depopulation lifetimes of Cs Rydberg states due to spontaneous emission and blackbody induced transitions. Red, green, blue curves are $T=300,77,4~\rm K$ respectively. Cs $ns_{1/2}, np_{3/2}, nd_{5/2}$ states are shown as dashed, solid, and dotted curves. The curves were calculated using the expressions in \cite{Beterov2009,Beterov2009b}. }
\label{fig.tau}
\end{figure}

Since all Rydberg gate protocols involve populating Rydberg states that have finite lifetimes there are principle limits to how high the fidelity can be. It was shown in \cite{Wesenberg2007} that in order to create one unit of entanglement between two-qubits with a dipolar interaction the integrated excitation under the gate operation must satisfy 
\begin{equation}
\int dt\, P_{\rm exc} \ge \frac{2}{V}
\label{eq.Elimit}\end{equation}
where $P_{\rm exc}$ is the two-atom excited state population and $V$ is the interaction strength. In the Rydberg context this implies that the gate error is bounded below by 
\begin{equation}
E\ge \frac{2}{{\sf B}\tau}
\end{equation} 
for a blockade interaction of strength ${\sf B}$.  The minimum error  of the blockade gate given by Eq. (\ref{eq.Emin}) has a $1/({\sf B}\tau)^{2/3}$ scaling and does not saturate the error bound. Likewise the interaction version of a Rydberg gate\cite{Jaksch2000} which works in the limit of 
 $\omega_q \gg \Omega\gg  V_{\rm dd}\gg 1/\tau$, with $V_{\rm dd}$ the weak dipolar interaction strength, has a minimum error of\cite{Saffman2010}
\begin{equation}
E_{\rm min} =\frac{\pi}{V_{\rm dd}\tau}+\frac{5 V_{\rm dd}}{3^{1/2}\omega_q}.
\label{eq.Emin2}
\end{equation}
Again the limit set by (\ref{eq.Elimit}) is not saturated. The fact that neither the blockade nor the interaction form of the Rydberg gate saturates the error bound suggests that it should be possible to devise a better protocol that does saturate the bound. This remains a challenge for future work.

Irrespective of whether or not the scaling of Eq. (\ref{eq.Elimit}) is achieved a set of parameters that lead to a high fidelity entangling gate is necessary for scalability. 
Despite the large body of work cited above it has been challenging to design a  gate protocol that can create Bell states with fidelity $F=0.9999$ when accounting for the parameters of real atoms. The difficulty lies in the conflicting requirements of running the gate fast enough to avoid spontaneous emission errors, yet slow enough to avoid blockade leakage errors. The requirement on gate speed can be clarified by looking at the spontaneous emission error in isolation. In order to create a Bell state with the blockade gate we start with the state $\ket{\psi_{\rm in}}=\frac{1}{\sqrt2}\left(\ket{01}+\ket{11}\right)$. Application of a CNOT gate gives the Bell state $\ket{\psi_{\rm out}}=\frac{1}{\sqrt2}\left(\ket{01}+\ket{10}\right)$. The time integrated Rydberg population during the CNOT is $P=7 t_\pi/4$ with $t_\pi=\pi/\Omega.$ Setting a limit $\epsilon_\tau$ on the spontaneous emission error implies 
$$
\tau> \frac{7}{4}\frac{t_\pi}{\epsilon_\tau}.
$$ 
The pulse time cannot be arbitrarily short due to the need to manage blockade leakage as well as practical considerations of available laser power and optical modulator bandwidth. For the sake of illustration let's put $t_\pi=25~\rm ns$ corresponding to a gate time of $4t_\pi=100~\rm ns.$ Figure \ref{fig.tau} shows the Rydberg lifetime as a function of principal quantum number together with the thresholds needed to reach $\epsilon_\tau=10^{-4},2\times 10^{-5}$.   We see that in a 300 K environment $\epsilon_\tau=10^{-4}$ is feasible for $n<120$ whereas  $\epsilon_\tau=2\times 10^{-5}$ or lower will only be practical in a 4 K environment. 

This shows that a recipe to reach $F=0.9999$ is a  $\sim 100~\rm ns$ gate time with a pulse sequence that strongly suppresses blockade errors. Since blockade errors arise at a sparse set of narrow leakage frequencies it is possible to design pulses that cancel 
leakage for such situations\cite{Theis2016}. Derivative removal by adiabatic gate (DRAG) has been highly successful at solving this problem for superconducting qubits\cite{Motzoi2009} and recent work has resulted in design of a Rydberg-DRAG gate with $F>0.9999$
for one-photon excitation of Cs atoms in a room temperature environment\cite{Theis2016b}. In principle similar fidelity could be obtained for two-photon excitation provided sufficient laser power is available for fast excitation at large detuning from the intermediate level to suppress spontaneous emission.

\subsubsection{Rydberg dressing}
Another class of related gate protocols relies on Rydberg dressing, whereby off-resonant Rydberg excitation admixes a small fraction of the Rydberg state into the ground state wavefunction thereby giving an effective interaction between atoms that largely stay in the ground state. This idea was originally introduced in the context of quantum gases\cite{Santos2000} and was  used recently to demonstrate a two-qubit entangling operation\cite{Jau2016}.

We can understand the fidelity limit of the dressing gate from a scaling analysis analogous to that used to analyze the blockade gate\cite{Saffman2005a,Saffman2010}. The effective ground state interaction in the dressing limit of $\Omega\ll \Delta$ is $V=-\Omega^4/8\Delta^3$, where $\Delta$ is the detuning of the dressing laser\cite{Johnson2010}. This limit is not entirely representative since the experiment \cite{Jau2016} was performed at an intermediate detuning with $\Omega\sim\Delta$, but allows us to make some analytical estimates.

The time to acquire  a two-atom conditional $\pi$ phase shift is 
$t_\pi = \pi/V$ giving a spontaneous emission error of $\epsilon_\tau=2\frac{\Omega^2}{2\Delta^2}t_\pi/\tau.$ The blockade leakage error is 
 $\epsilon_\bshift=2\frac{\Omega^2}{2\Delta^2}.$ Choosing $\Omega$ to minimize the sum of the errors we find 
\begin{equation}
E_{\rm min}=\frac{2^{5/2}\pi^{1/2}}{(\Delta\tau)^{1/2}}.
\end{equation}
As with the blockade and interaction gates we do not saturate the scaling of Eq. (\ref{eq.Elimit}). 
The maximum detuning at large $n$ is $\hbar\Delta=\delta U/2\sim E_{\rm H}/(2 n^3)$
and again using $\tau=\tau_0 n^3$ we find $E_{\rm min}\sim 8\pi^{1/2}(\hbar/E_{\rm H}\tau_0)^{1/2}.$ Putting in numerical values we find an
error floor $E_{\rm min}=0.0013$. Numerical analysis that accounts for both spontaneous emission and finite Rydberg interaction strength  predicts a somewhat higher error floor of $0.002$ after averaging over product states in the computational basis\cite{Keating2015}. 
This error is comparable to that found for Bell state preparation for the blockade gate with constant amplitude laser pulses\cite{XZhang2012}.

\subsubsection{Dissipative entanglement} An alternative to coherent evolution is to create entangled states by a combination of coherent and dissipative driving. If the dark state of the dissipative dynamics is an entangled state we may obtain high fidelity entanglement despite a high level of spontaneous emission\cite{Plenio1999,Diehl2008,Verstraete2009,Reiter2015}. Entanglement of two trapped ion qubits has been demonstrated in this way\cite{YLin2013}. Dissipative dynamics together with Rydberg interactions were first considered for creating multiparticle correlations and entanglement\cite{Weimer2010,Lee2011},  and later extended to preparation of two-qubit Bell states\cite{Carr2013,Rao2013}. Subsequent work has studied simplified  protocols\cite{Su2015}, extensions to higher dimensional entanglement\cite{Shao2014},  as well as  
manybody dynamics and spin correlations\cite{Hu2013,Otterbach2014,Schonleber2014,Hoening2014,Rao2014b,Weimer2015b,Lee2015,Morigi2015,Ray2016}. Experimental signatures of the role of dissipation in Rydberg excitation statistics were reported in \cite{Malossi2014}.

Since spontaneous emission does not constitute an error for dissipative 
gate protocols it might be hoped that very high fidelity entanglement could be obtained. Although the scaling is different than for blockade gates, also for dissipative dynamics the fidelity cannot be arbitrarily high in real atoms. The limiting factors vary depending on the details of the protocol used  but the fidelity limit can be understood as being due to the fact that for any finite rate of entanglement generation the entangled dark states are not perfectly dark. The fidelity of the target state is  determined by a balance between the pumping rate into  and depumping rate out  of the dark state. Detailed analyses have shown maximal Bell state fidelities of 0.998\cite{Carr2013} and 0.995\cite{Rao2013}. Nevertheless, since dissipative methods require much reduced Rydberg interaction strengths they may be useful for creating entanglement at long range, beyond the reach of a blockade gate. The imperfect entanglement so created could then in principle be purified if higher fidelity local operations are available\cite{Bennett1996p,Bennett1996perratum}.

\subsection{Multi-qubit gates}
\label{sec.gatesm}

One of the attractive features of Rydberg interactions is that multi-particle entanglement and logic operations can be generated efficiently. Although any multi-qubit operator can be decomposed into a sequence of one- and two-qubit gates the decomposition is often inefficient. The long range nature of the Rydberg interaction whereby a single Rydberg excited qubit can block the excitation of $k$ target qubits serves as a primitive for a CNOT$^k$ gate. This idea originated in \cite{Lukin2001}, was further analyzed in \cite{Unanyan2002}, and has been used to create entangled $\ket{W}$ states\cite{Zeiher2015,Ebert2015} for encoding of  ensemble qubits. The Rydberg blockade strength has also been experimentally characterized in detail and shown to agree with ab-initio calculations for three interacting atoms\cite{Barredo2014}.

A related interaction mechanism can be used for one step generation of GHZ states\cite{Brion2007c,Saffman2009b,Muller2009,Opatrny2012} and has been proposed for implementation of topological spin models\cite{Weimer2010,Weimer2011}. Adiabatic protocols for creating multiparticle entanglement have been analyzed in \cite{Cano2014,RCYang2016}.

The complementary situation where the joint state of $k$ control qubits  acts on a single target qubit can be used for implementing a C$_k$NOT gate\cite{Isenhower2011}. When $k=2$ this corresponds to the three qubit Toffoli gate that finds frequent use in error correcting codes. For larger $k$ the C$_k$NOT is a primitive for the quantum search algorithm\cite{Grover1997} whereby Rydberg interactions enable highly efficient implementation of quantum search\cite{Molmer2011,Petrosyan2016}. Although the fidelity of the C$_k$NOT gate decreases with increasing $k$ quantum Monte Carlo simulations have shown that for $k=2$ the native C$_2$NOT Rydberg gate has better fidelity than decomposition into single and two-qubit Rydberg gates\cite{Gulliksen2015}.

\subsection{Experimental issues for Rydberg gates}
\label{sec.experimental}

Even with an ideal two-qubit interaction Hamiltonian experimental imperfections cause errors that reduce fidelity.  The Bell state fidelity listed in Table \ref{tab.gate} includes state preparation and measurement errors. Also imperfect experimental control reduces gate fidelity below the intrinsic theoretical limit. The dominant experimental issues for Rydberg gates are finite temperature Doppler dephasing, laser noise and pointing stability,
gate phases due to Stark shifts, spontaneous emission from the intermediate state in two-photon Rydberg excitation, and perturbations to Rydberg states due to background electric and magnetic fields. 
 Many of these issues have been reviewed previously\cite{Saffman2005a,Saffman2010} as well as in a recent technical guide\cite{Naber2015b}. Here we give a brief update with an eye to what will be needed to reach a fidelity goal of $F=0.9999$. 

Atoms that are not cooled to the vibrational ground state of the trapping potential exhibit Doppler detuning upon excitation to Rydberg states. This leads to dephasing and reduced Bell state fidelity as was first pointed out in \cite{Wilk2010}. The Doppler limited Bell state fidelity is\cite{Saffman2011}
\begin{equation} 
F_D=\frac{1+e^{-\frac{k^2 k_B T t^2}{2 m}}}{2}
\label{eq.FDoppler}
\end{equation}
with $m$ the atomic mass, $k$ the magnitude of the Rydberg excitation wavevector, $T$ the atomic temperature, and $t$ the time spent in the Rydberg state. The infidelity $1-F_D$ should be added to the intrinsic infidelity discussed in Sec. \ref{sec.intrinsic}. 

\begin{figure}[!t]
\centering
\begin{minipage}[c]{8.cm}
\centering
 \includegraphics[width=8.cm]{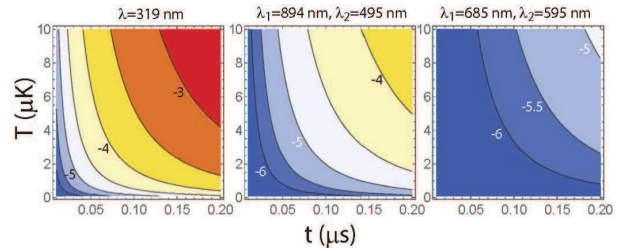}
\end{minipage}
  \caption{Doppler limited Bell fidelity from Eq. (\ref{eq.FDoppler}) for Cs atoms and three different excitation methods: left) One photon, center)
two counterpropagating photons via the $6p_{1/2}$ state, and right) two counterpropagating photons via the $5d_{5/2}$ state\cite{Xia2012}. The contours are labeled with $\log_{10}(1-F_D)$.}
\label{fig.FDoppler}
\end{figure}

Figure \ref{fig.FDoppler} shows the Doppler limited Bell fidelity as a function of $k$ and atom temperature for three different excitation methods. The one photon excitation case\cite{Jau2016} in Fig. \ref{fig.FDoppler} has the highest Doppler sensitivity although it has the advantage that there is no spontaneous emission from an intermediate state. Two-qubit blockade via one-photon excitation of Cs $np_{3/2}$ states was demonstrated in \cite{Hankin2014}. We see that the requirement on gate time to reach $F_D=0.9999$ is similar to the requirement set by spontaneous emission in Fig. \ref{fig.tau} provided the atoms are cooled to $<5 ~\mu\rm K$. The availability of high power 319 nm single frequency radiation renders this a viable approach to high fidelity gates\cite{Rengelink2016, JWang2016}. The two-photon excitation cases substantially relax the cooling requirement to reach   $F_D=0.9999$, although performing a two-photon excitation with low spontaneous emission at sub 100 ns timescales puts severe requirements on laser intensity.  Also three photon Rydberg excitation can be used in which case it is possible to eliminate Doppler broadening\cite{Ryabtsev2011}.

In addition to Doppler errors laser amplitude, phase, and frequency noise, as well as the spatial profile and beam pointing all impact gate fidelity. For single qubit gate operations composite pulse sequences can be used to reduce sensitivity to technical fluctuations. Unfortunately composite pulses trade reduced sensitivity to noise against longer gate times and are therefore not as useful for Rydberg gates as for ground state operations. Optimized beam shaping can reduce sensitivity to pointing errors without invoking longer gate times\cite{Gillen-Christandl2016}.  

Another issue relevant to Rydberg gates is the presence of differential dynamic Stark shifts between the ground and Rydberg states. The blockade gate protocol\cite{Jaksch2000} is based on the appearance of a $\pi$ phase shift after a resonant Rabi  pulse with area $2\pi$. When multi-photon excitation is used a $2\pi$ resonant pulse will in general give a wavefunction phase different from $\pi$ that depends on AC Stark shifts arising from the excitation fields. This necessitates tuning of parameters to recover an ideal entangling gate.   A detailed discussion of these issues can be found in\cite{Maller2015}.  

The presence of hyperfine substructure, although usually ignored in analysis of Rydberg gates, may impact gate performance. Particularly in Cs which has large hyperfine splittings\cite{Sassmannshausen2013}  partially resolved excitation of multiple hyperfine Zeeman states can lead to significant changes in gate performance. Such issues can be overcome by using selection rules to only allow excitation of a single hyperfine state. For example, starting in the Cs stretched ground state $\ket{6s_{1/2},f=4,m_f=4}$ and applying a $\sigma_+$ polarized excitation field will only couple to $\ket{np_{3/2},f=5, m_f=5}$. Also two-photon excitation can be designed to couple to a single hyperfine state. An example being $\sigma_+,\sigma_+$ excitation of $\ket{6s_{1/2},f=4,m_f=4}$ to $\ket{6p_{3/2},f=5,m_f=5}$ to $\ket{nd_{5/2},f=6,m_f=6}$. Imperfect polarization of the excitation lasers will allow coupling to other states so good polarization purity is a requirement for high fidelity control. 

Finally there is the issue of Rydberg sensitivity to background  electric fields. Robust control of multi-qubit experiments requires stable optical addressing which is enabled by designing miniaturized geometries, possibly based on atom chip technology. Hybrid quantum interfaces with Rydberg atoms also involve near surface 
geometries\cite{Petrosyan2009,Patton2013,Pritchard2014}. 
Rydberg atoms  subjected to fields from surface charges can be strongly perturbed  since the dc polarizability of Rydberg states scales as $\alpha_{\rm Ryd, dc}\sim n^7$. Also in trapped ion approaches to quantum computing  surface fields are of major concern which has motivated detailed studies of field noise from surfaces\cite{Brownnutt2015}.

 Several groups have measured and characterized near surface  electric fields using methods such as the motion of atoms in a Bose-Einstein condensate \cite{Obrecht2007}, motional spectroscopy of trapped ions \cite{Brownnutt2015}, Rydberg electromagnetically induced transparency \cite{Tauschinsky2010,Abel2011,Hattermann2012,Chan2014}, and Rydberg Stark spectroscopy \cite{Carter2011,Carter2012,Naber2016,Thiele2015}.
Fields with magnitudes of $0.1-10~\rm V/cm$
have been measured at distances of 10-100 $\mu\rm m$.  If the fields are static, and not too large, then the Rydberg excitation laser can be tuned to account for the Stark shift from the background field. Time varying fields would lead to fluctuating detuning and gate errors. Even static fields can be problematic if they are large enough to allow detrimental couplings to otherwise forbidden states. 

To put the problem in perspective consider a Cs atom in the Rydberg state $100p_{3/2}$. The dc scalar and tensor polarizabilities are  
$\alpha_0=205 ~\rm GHz/(V/cm)^2,$ $\alpha_2=-17.8 ~\rm GHz/(V/cm)^2$. A high fidelity Rydberg gate with duration 100 ns requires excitation 
Rabi frequencies of $\Omega/2\pi = 20~\rm MHz$. A $10^{-5}$ error in the Rydberg excitation probability after a $\pi$ pulse requires a detuning error of not more than about 90 kHz. This translates into a field limit of $\delta E<6.6\times 10^{-4} ~\rm V/cm. $ This field strength is several orders of magnitude smaller than the fields that have been measured within 100 $\mu\rm m$ from surfaces. Even in cm scale vacuum cells background fields $>0.02 ~\rm V/cm$ are routinely observed. 

There are several routes to mitigation of this problem. Electrodes can be placed inside the vacuum cell for dc field 
cancellation\cite{Hofmann2014}. It has also been shown to be possible to suppress the development of background fields by purposefully coating proximal surfaces with a  layer of alkali adsorbates. On a metallic  chip surface this provides a  uniform conducting layer that prevents additional adsorbates accumulating\cite{Hermann-Avigliano2014}.
The effect on a quartz substrate is to induce negative electron affinity which binds low energy electrons, canceling the field from the alkali adsorbates\cite{Sedlacek2016}. Surface baking to uniformly diffuse adsorbates has also been shown to have a beneficial 
effect\cite{Obrecht2007}. Substantial reduction in field strength can be achieved. For example in the experiments with a quartz substrate fields of only $0.03~\rm V/cm$ were observed at $20~\mu\rm m$ from the surface. 

Further reduction of Rydberg perturbations from  dc fields can be
achieved by admixing Rydberg states with opposite sign of polarizability using microwave fields\cite{Mozley2005,Jones2013}. In \cite{Jones2013}, microwave fields at \numunit{\sim\nobreak 38}{GHz} are used to cancel the relative polarizabilities between the $\state{48s}{1/2}$ and $\state{49s}{1/2}$ Rydberg levels in $^{87}\mathrm{Rb}$, coupling the $s$ states to neighboring $p$ states. Although the $p$ states have polarizabilities of the same sign as $s$ states and thus cannot cancel the absolute $s$ state polarizabilities, the experiment aimed to cancel the relative Stark shift between two Rydberg levels. In \cite{Ni2015}, the Martin group expanded this relative polarizability cancellation to pairs of circular Rydberg states. 
Although more work remains to be done it is likely that a combination of careful attention to surface preparation and Rydberg dressing for polarizability cancellation will enable high fidelity Rydberg control at atom-surface distances of a few tens of microns.

Finally there is the question of scalability of high fidelity gates to large multi-qubit systems. High gate fidelity puts requirements on optical power to maintain large detuning from intermediate excited states for Stark shifted one-qubit gates, and to achieve fast Rydberg excitation for entangling gates.   Global single qubit gates  performed with microwaves, as in \cite{Xia2015}, easily scale to very large numbers of qubits. Site selected gates relying on focused optical beams, as discussed in Sec. \ref{sec.array},  imply a laser power requirement that scales proportional to $p$, the number of gates performed in parallel during a single time step. 
Scalable quantum computing requires error correction which in turn implies that $p$ must necessarily grow  with  the total number of qubits $N$. If this were not the case then the rate at which error correction could be applied to a logical qubit would decrease with system size and the correction would eventually fail. The required value of $p/N$ will depend on the details of the error correction code used, but we expect the ratio to be roughly constant as $N$ increases.

We see that a scalable laser power resource is a requirement for qubit number scalability. This is a technical and economic challenge, not a fundamental one.  It may be mentioned that extremely low noise  laser sources with power greater than 100 W have already been developed for scientific projects such as gravitational wave detection\cite{Kwee2012}. Another solution is to deploy arrays of low to moderate power lasers with each qubit, or group of qubits, being controlled by their own lasers\cite{Burd2016}. As current experiments are far from the regime where laser power is the main limitation this issue is not yet of primary importance  but will eventually have to be tackled in order to engineer large scale systems.  

\section{Other approaches}
\label{sec.other}

We have so far concentrated on the use of single alkali atom qubits for circuit model, gate based quantum computation. This is indeed the most advanced neutral atom approach to quantum computing at this time.
However, there are also other possibilities for qubit encoding and for computation. As noted in Fig. \ref{fig.encoding} ensembles can be used for encoding of single qubits or collective encoding of qubit registers. Ensembles are also of interest for mediating atom-light entanglement\cite{LiDudin2013} and quantum networking. One of the challenges of ensemble qubits is the $\sqrt N$ scaling  of the excitation Rabi frequency with the number of qubits $N$ in the ensemble. Although this scaling is a hallmark of blockade physics it makes the execution of high fidelity gate operations problematic when $N$ is not accurately known. Adiabatic gate protocols have been proposed to suppress the dependence on $N$ and a universal set of ensemble gates can be constructed\cite{Beterov2011,Beterov2013a,Beterov2014,Beterov2016}. 

In addition to standard gate operations Rydberg interactions are of interest for encoding logical qubits in decoherence free 
subspaces\cite{Brion2007a}, for coherent versions of quantum error correction\cite{Crow2015}, 
 and preparation of topological states\cite{Weimer2010,Weimer2011,Nielsen2010b,Lesanovsky2012}.
Beyond circuit model  computation Rydberg interactions have been considered for alternative paradigms including one-way quantum computing\cite{Kuznetsova2012} and adiabatic quantum computing\cite{Keating2013}. 

\subsection{Quantum simulation and dressing}
\label{sec.dressing}

More generally there is great potential for using Rydberg interactions for quantum simulation\cite{Cirac2012,Bloch2012,Hauke2012}. The availability of long range dipolar interactions  with anisotropy that can be controlled by clever choice of the interacting Rydberg states allows for study of many-body interactions and tailored spin models\cite{Pohl2010,Ji2011,Hague2012,Hague2014,Cesa2013,Glaetzle2014,Vermersch2015,Glaetzle2015,JQIan2015}. Also Rydberg dressing\cite{Santos2000,Henkel2010,Honer2010,Johnson2010,Balewski2014,Macri2014} can be used to engineer a remarkably wide variety of interaction Hamiltonians and phenomena\cite{Pupillo2010,Cinti2010,Maucher2011,
Henkel2012,Glaetzle2012, Dauphin2012,Hsueh2012,
Mattioli2013,Grusdt2013,Hsueh2013,Mobius2013,
Xiong2014,
vanBijnen2015b,Lan2015,Levi2015,Dalmonte2015,XLi2015,
Angelone2016,Dauphin2016,Chougale2016}. 
 Paradigmatic experimental advances have demonstrated signatures of long range  and spin dependent interactions\cite{Schauss2012,Barredo2015,Schauss2015,Zeiher2016,Labuhn2016}, and this remains a rich area for future work.

\begin{figure}[!t]
\centering
\begin{minipage}[c]{8.cm}
\centering
 \includegraphics[width=8.cm]{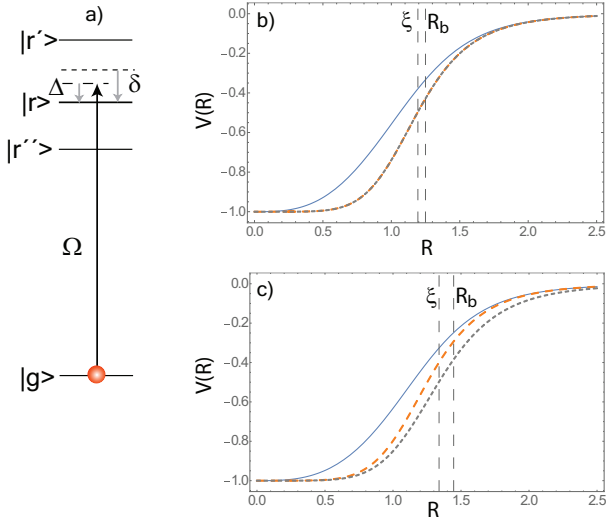}
\end{minipage}
  \caption{a) Dressing of ground state atoms by off-resonant excitation of Rydberg state $\ket{r}$ with Rabi frequency $\Omega$ and detuning $\Delta$. Opposite parity Rydberg levels $\ket{r'}, \ket{r''}$ mediate the dipole-dipole interaction with F\"orster 
defect $\delta$.  b) The normalized interaction strength
$V(R)=[\Delta_{\rm dr}(R)-\Delta_{\rm dr}(\infty)]/|\Delta_{\rm dr}(0)-\Delta_{\rm dr}(\infty)|$ with $\Delta_{\rm dr}$ given by Eqs. (\ref{eq.Vdd},\ref{eq.Vdr}) (solid blue line), $\Delta_{\rm dr}$ given by Eqs. (\ref{eq.VvdW},\ref{eq.Vdr}) (dashed orange line), and using the van der Waals approximation\cite{Honer2010} $V(R) = -\xi^6/(R^6 + \xi^6) $ with 
$\xi=R_c \left(\frac{\delta}{8\Delta}\right)^{1/6}$ (dotted gray line). When $\delta=\Delta$ then $\xi=R_b$ from Eq. (\ref{eq.Rb}). Parameters are  $\Omega/2\pi = 1,$ $\Delta/2\pi = 10,$ $\delta/2\pi=20$, and $R_c=1.5$. c) Same as b) except $\Omega/2\pi = 10,$ $\Delta/2\pi = 5.$ }
\label{fig.dressing}
\end{figure}

As regards Rydberg dressing the ultimate potential  for high fidelity implementation of dressing Hamiltonians is somewhat unclear at this time due to higher decoherence rates than would be expected from a naive analysis
\cite{Aman2016,Goldschmidt2016}. To understand the issues involved consider a ground state that is coupled to Rydberg state $\ket{r}$ with Rabi frequency $\Omega$ and detuning $\Delta$ as shown in Fig. \ref{fig.dressing}. The dipole-dipole interaction of two atoms in Rydberg $\ket{ns_{1/2}}$ states at separation $R$ can be expressed as a frequency shift\cite{Walker2008}
\begin{equation}
\Delta_{\rm dd}(R)=\frac{\delta}{2}-\frac{\delta}{2}\left[1+(R_c/R)^6 \right]^{1/2}.
\label{eq.Vdd}
\end{equation}
Here $\delta$ is the F\"orster defect of the pair interaction, $R_c=\left(\frac{4 D_{kl} C_3^2}{\hbar^2\delta^2} \right)^{1/6}$ is the crossover length scale between resonant dipole-dipole and van der Waals regimes and $C_3$, which is proportional to $n^4$ and radial matrix elements between Rydberg states, determines the strength of the interaction.  The angular factor $D_{kl}$ depends on the Zeeman substates of the atoms and the angular momenta of the interaction channel\cite{Walker2008,Beterov2015}. For the specific case of alkali atoms in a triplet spin state with $M=\pm 1$ and   coupling of initial $ns_{1/2}$ states to  $np_{1/2,3/2}$ states, averaged over the fine structure,  $D_{kl}=12$ using the definitions of Ref.\cite{Beterov2015}. For $R\gg R_c$ the interaction (\ref{eq.Vdd}) leads to a long range van der Waals interaction
\begin{equation}
\Delta_{\rm vdW}(R)=-\frac{\delta}{4}(R_c/R)^6.
\label{eq.VvdW}
\end{equation}

The sign of the interaction depends on the sign of $\delta$ which is determined by the quantum defects. For $\delta>0(<0)$ we find $d\Delta_{\rm dd}/dR>0(<0)$ and an attractive(repulsive) interaction.  The combination of $\delta>0$  and $\Delta<0$ leads to an excitation  resonance at finite $R$ as does the combination $\delta<0$  and $\Delta>0$. We will exclude these cases and only consider the combinations of $ \delta>0,\Delta>0 $ giving attractive potentials and $ \delta<0,\Delta<0 $ giving repulsive potentials.  

At small separation the interaction strength is modified by blockade physics since only one atom at a time can be Rydberg excited. This results in a soft core potential.  Far off-resonance dressing uses  $|\Delta|\gg |\Omega|$ and in this regime we define a blockade distance $R_b$ by 
$|\Delta_{\rm dd}(R_b)|=|\Delta|$. In the weak excitation limit $R_b$ corresponds to the distance at which the excitation probability is suppressed by a factor of 4. Solving Eq. (\ref{eq.Vdd}) we find 
\begin{equation}
R_b=R_c\frac{|\delta|^{1/3}}{2^{1/3}[\Delta(\Delta+\delta)]^{1/6}}.
\label{eq.Rb}
\end{equation}
For $50<n<100$ in the heavy alkalis $\delta/2\pi$ is in the range of about $0.2 - 2 ~\rm GHz$\cite{Beterov2015}. Taking for example $\Delta=\delta$ we find $R_b=R_c/\sqrt2$.

In the symmetric basis $\left\{\ket{gg},\frac{\ket{gr}+\ket{rg}}{\sqrt2},\ket{rr}\right\}$ the two-atom Hamiltonian is 
$$
H = \hbar\left(\begin{array}{ccc}
0 & \Omega^*/\sqrt2& 0\\
\Omega/\sqrt2& -\Delta& \Omega^*/\sqrt2\\
0 & \Omega/\sqrt2 &  -2\Delta+\Delta_{\rm dd}
\end{array}\right).
$$
%$$
%H = \hbar\begin{pmatrix}
%0 & \Omega^*/\sqrt2& 0\\
%\Omega/\sqrt2& -\Delta& \Omega^*/\sqrt2\\
%0 & \Omega/\sqrt2 &  -2\Delta+\Delta_{\rm dd}
%\end{pmatrix}.
%$$
Solving for the eigenvalues  the energy of the dressed ground state is given by
\begin{eqnarray}
\Delta_{\rm dr}(R)&=& -\Delta+\frac{\Delta_{\rm dd}}{3}\nonumber\\
&+&\frac{2^{2/3}\left(\Delta^2- \Delta\Delta_{\rm dd} + \frac{1}{3}\Delta_{\rm dd} ^2 +|\Omega|^2\right)}{f}+\frac{2^{1/3}}{6}f\nonumber\\
\label{eq.Vdr}
\end{eqnarray}
with 
%%\begin{widetext}
\begin{eqnarray}f&=&\left[18\Delta\Delta_{\rm dd}(\Delta-\Delta_{\rm dd})+4\Delta_{\rm dd}^3-9\Delta_{\rm dd}|\Omega|^2\right.\nonumber\\
&&\left. +\left[\Delta_{\rm dd}^2\left(18\Delta^2-18\Delta \Delta_{\rm dd}+4 \Delta_{\rm dd}^2-9|\Omega|^2 \right)^2 \right.\right.\nonumber\\
&&\left.\left.-16\left(3\Delta^2-3 \Delta \Delta_{\rm dd}+\Delta_{\rm dd}^2+3|\Omega|^2 \right)^3\right]^{1/2} \right]^{1/3}.
\end{eqnarray}
%%\end{widetext}
Using $\Delta_{\rm dd}=\Delta_{\rm dd}(R)$ from (\ref{eq.Vdd}) we obtain an  exact, albeit cumbersome,  expression for $\Delta_{\rm dr}(R)
$ in the limit of a single Rydberg interaction channel. Equation (\ref{eq.Vdr}) generalizes the widely used approximation of $\Delta_{\rm dr}\sim\left[1+(R/R_c)^6\right]^{-1}$ \cite{Henkel2010,Honer2010} which assumes a pure $1/R^6$ 
van der Waals interaction.

For $R\gg R_c$  the effective ground state potential is just the light shift of two noninteracting atoms $\Delta_{\rm dr}(\infty) = -\Delta+\sqrt{\Delta^2+|\Omega|^2}$. For $R\ll R_c$  we get  a one atom light shift of the blockaded two-atom state $\Delta_{\rm dr}(0) = -\frac{\Delta}{2}+\frac{\sqrt{\Delta^2+2|\Omega|^2}}{2}$. The depth of the effective potential is thus 
 the difference between the one- and two-atom light shifts\cite{Johnson2010}
\begin{eqnarray}
\Delta_{\rm dr}(0)-\Delta_{\rm dr}(\infty)&=& \frac{\Delta}{2} + \frac{\sqrt{\Delta^2 + 2 |\Omega|^2}}{2}- \sqrt{\Delta^2 +  |\Omega|^2}\nonumber\\
&\simeq& - \frac{|\Omega|^4}{8\Delta^3}.\nonumber
\end{eqnarray}
 We emphasize that the above expressions only describe a two-atom interaction. In a dense gas with $|\Omega/\Delta|$ not sufficiently small the dressing will take on the character of a collective many body interaction. Expressions for the cross over conditions   between two body and collective effects  effects can be found in \cite{Honer2010,Balewski2014}. 

Figure \ref{fig.dressing} compares  exact and approximate forms of the normalized soft-core potentials.  Using either the full dipole-dipole interaction or the van der Waals approximation there is a smooth soft core potential but the shape at small $R$ is quite different. The leading term near the origin 
is proportional to $R^3$ using Eq. (\ref{eq.Vdd}) and $\sim R^6$ using Eq. (\ref{eq.VvdW}). Thus the vdW approximation shows a flatter core potential than the real dipolar interaction. The difference between the full expression (\ref{eq.Vdr}) with the van der Waals potential (\ref{eq.VvdW}) and the single term van der Waals approximation \cite{Henkel2010,Honer2010} is negligible in the limit of weak dressing in Fig. \ref{fig.dressing}b), while for $|\Omega|$ compaable to $|\Delta|$ in Fig. \ref{fig.dressing}c)  a substantial difference is seen between Eqs.  (\ref{eq.VvdW},\ref{eq.Vdr})  and the single term approximation.  Very good agreement between measured and calculated dressing curves was obtained in Ref.\cite{Jau2016}. Unfortunately those measurements cannot be directly compared with  single channel theory since the data was taken in a regime where multiple  channels contributed and the calculations involved accounting numerically for the contribution from a large number of Rydberg levels.

With this description of the dressing interaction in hand we can define a figure of merit for implementation of many-body dressing Hamiltonians.  The characteristic decoherence time scale per atom due to spontaneous and blackbody induced Rydberg depopulation is 
$$
\tau_{\rm dr} = \frac{2\Delta^2}{|\Omega|^2}\tau
$$
with $\tau$ the Rydberg state lifetime.  The number of coherent operations due to the dressing interaction in one decoherence time is  $\sim|\Delta_{\rm dr}(0)-\Delta_{\rm dr}(\infty)| \tau_{\rm dr}/2\pi$ and multiplying by $N_{\rm 3D}=\frac{4\pi}{3}(R_b/2d)^3$, the number of atoms in  a qubit lattice with period $d$ inside a spherical volume of diameter $R_b$, 
we define a figure of merit 
\begin{eqnarray}
F_{\rm 3D}&=&  \frac{1}{2\pi}|\Delta_{\rm dr}(0)-\Delta_{\rm dr}(\infty) |\tau_{\rm dr}N_{\rm 3D}\nonumber\\
&\simeq&\frac{1 }{96}\frac{|\Omega|^2|\delta|}{|\Delta|^{3/2}|\Delta+\delta|^{1/2}}\tau \left(\frac{R_c}{d}\right)^3.\label{eq.dressing}\nonumber
\end{eqnarray}
The corresponding expressions in lower dimensions are 
\begin{eqnarray}
F_{\rm 1D}&\simeq&\frac{1 }{ 2^{1/3}8\pi}\frac{|\Omega|^2|\delta|^{1/3}}{|\Delta|^{7/6}|\Delta+\delta|^{1/6}}\tau \left(\frac{R_c}{d}\right), \nonumber\\
F_{\rm 2D}&\simeq&\frac{1 }{2^{2/3}32}\frac{|\Omega|^2|\delta|^{2/3}}{|\Delta|^{4/3}|\Delta+\delta|^{1/3}}\tau \left(\frac{R_c}{d}\right)^2 . \nonumber
\end{eqnarray}

The scalings with principal quantum number in the heavy alkalis are $\delta\sim {1/n^4}$, $\tau\sim n^3$ and $R_c\sim n^{8/3}$. In order to avoid overlap of the Rydberg electron wavefunction with a neighboring ground state atom we require $d\stackrel{>}{\sim} a_0 n^2$ and in order to couple primarily to a single $n$ state we require $\Delta\sim 1/n^3$. We find  
asymptotically $F_{\rm 1D, 2D, 3D}\sim n^{19/3}, n^{20/3}, n^{7}$. 
In all cases the figure of merit, which corresponds to the number of coherent operations times the number of interacting atoms, scales as a high power of $n$. 

To get  a sense of what is possible consider ground state   Cs atoms dressed by  $\ket{ns,ns,M=1}$ pair states and a dipole-dipole interaction from the channel coupling to $\ket{np_{3/2}, (n-1)p_{3/2}, M=1}$. At  $n=100$ we have  $\delta/2\pi=-200.~\rm MHz $, $R_c=8.1~\mu\rm m$, and $\tau=320~\mu\rm s $.  Although $np$ states have longer lifetimes, there are angular zeroes in the interaction\cite{Walker2008} so we have assumed $ns$ states. 
With $\Omega/2\pi = 20~\rm MHz$, $\Delta/2\pi=100~\rm MHz$, $d=1~\mu\rm m$ we find 
\begin{eqnarray}
F_{\rm 1D, 2D, 3D}&=&2200,~~11000,~~51000 ,\nonumber\\
N_{\rm 1D, 2D, 3D}&=& 6,~~35,~~160,\nonumber
\end{eqnarray}
with $|\Delta_{\rm dr}(0)-\Delta_{\rm dr}(\infty)|=2\pi\times 20~\rm kHz$,  $\tau_{\rm dr}=16~\rm ms$ and $F/N = 320$ operations per atom. 
 These estimates verify that interesting simulations of many-body coherent quantum dynamics should be possible using Rydberg dressing.

Nevertheless there has been recent concern about the viability of dressing for studying unitary evolution due to anomalously short decoherence times. 
Decay of a Rydberg excited atom at large $n$ in a room temperature environment will, with probability approximately 1/2, populate a neighboring opposite parity Rydberg level\cite{Low2012}. The presence of a single Rydberg atom in an opposite parity state leads to a resonant dipole-dipole interaction that rapidly leads to state transfer, and an avalanche depopulation  of the many-body Rydberg state. Signatures of this dynamics have been 
seen in Ref. \cite{Goldschmidt2016,Aman2016}.

Assuming that avalanche depopulation proceeds rapidly compared to the other time scales of the problem the effective 
 coherence time in a many atom sample is reduced by a factor of $N/2$. This gives a modified figure of merit 
 \begin{eqnarray}
F'&=&  \frac{2}{2\pi}|\Delta_{\rm dr}(0)-\Delta_{\rm dr}(\infty) |\tau_{\rm dr} \simeq\frac{|\Omega|^2}{4\pi|\delta|}\tau.\label{eq.dressing2}
\end{eqnarray}
The asymptotic scaling of this modified figure of merit is $F'\sim n^6$ independent of the dimensionality. 
Using the same parameters as in the previous paragraph we find $F'=640$
and $F/N_{\rm 1D, 2D,3D}=95,~ 18,~ 4$.
Since the right hand side of (\ref{eq.dressing2}) is independent of $N$ the available number of coherent evolution steps per atom is reduced as $N$ is increased, which is obvious when faced with a decoherence rate that scales with $N$. Nevertheless significant multistep coherent evolution appears possible, particularly  in lower dimensions. 
Further improvements may stem from increasing $\Omega$ or $n$ to reduce $|\delta|$, cryogenic environments to increase $\tau$, or  from new approaches such as a combination of electromagnetically induced transparency and dressing\cite{Gaul2016}.

\subsection{Other species}

While alkali atoms have received the most attention for Rydberg experiments due to their experimental simplicity other elements provide new opportunities. Alkaline earth elements have two $s$ electrons. One can be excited for Rydberg interactions while the other electron can provide a useful handle for trapping and cooling. 
Progress in Rydberg physics with alkaline earth atoms is reviewed in \cite{Dunning2016}.  The lanthanides Er, Dy, Th, and Ho have been the subject of rapidly increasing interest for ultracold experiments. Holmium was proposed for collective encoding\cite{Saffman2008} since it has the largest number of ground hyperfine states of any element. Trapping and precision Rydberg spectroscopy of Ho atoms\cite{Miao2014,Hostetter2015} have revealed regular Rydberg series despite the complexity of the electronic structure. The measured quantum defects suggest that strong Rydberg interactions should be observable in future experiments. Also the large ground and excited state hyperfine splittings due to the open $4f$ shell which imply high fidelity for optical pumping and state measurements may make the lanthanides competitive for single atom qubit encoding.

\subsection{Hybrid interfaces}

Hybrid quantum systems involving different matter based qubits are being developed in many different 
directions\cite{Xiang2013,Kurizki2015}.
The exceptional strength of Rydberg interactions is attractive for coupling to not only neutral atoms but also ions, molecules, optomechanical systems, and superconductors. Rydberg interactions have been proposed as an alternative to the usual Coulomb gates for 
trapped ions\cite{Muller2008,Schmidt-Kaler2011,WLi2012,WLi2013,WLi2014,Bachor2016}, for 
studying 2D spin models in ion crystals\cite{Nath2015}, and  for coupling ions to neutral atoms\cite{Secker2016}. Excitation of a trapped $^{40}$Ca$^+$ ion to a Rydberg state was demonstrated in \cite{Feldker2015} by preparation of the ion in a metastable low lying  level followed by a one-photon transition with a vacuum ultraviolet photon.  Rydberg mediated coupling between 
polar molecules and neutral atoms\cite{Rittenhouse2010,Zhao2012} has been proposed for molecular quantum gates\cite{Kuznetsova2011}, and  for readout of the rotational state of polar molecules\cite{Kuznetsova2016}. Optomechanical effects due to coupling of membranes to atomic Rydberg states have been discussed in\cite{Carmele2014,Bariani2014,DYan2015}. There are also active efforts to couple Rydberg atoms to quantum states of microwave 
photons\cite{Sorensen2004} as part of an interface between atomic and superconducting qubits\cite{Petrosyan2009,Hogan2012,Lancuba2013,Carter2013,Pritchard2014,Hermann-Avigliano2014,Thiele2014,
Sarkany2015,Thiele2015,Lancuba2016}.

\section{Outlook}
\label{sec.outlook}

We have taken a critical look at the potential of neutral atoms with Rydberg interactions for scalable quantum computation and simulation. 
Neutral atom approaches can conceivably provide many thousands of qubits in a small footprint of less than $1 ~\rm mm^2$, an extremely attractive capability that is  difficult to match with any other technology. In the  six years since  the first entanglement demonstrations\cite{Wilk2010,Isenhower2010} there has been palpable experimental progress towards larger qubit arrays\cite{Nogrette2014,Xia2015,YWang2016}, deterministic atom loading\cite{Endres2016,Barredo2016}, higher fidelity entanglement\cite{Maller2015,Jau2016}, and preparation of ensemble qubits\cite{Ebert2015,Zeiher2015}. Quantum simulation, in particular using Rydberg dressing, has attracted a great deal of interest and there are very promising 
results in both theory\cite{Glaetzle2014,vanBijnen2015b,Lan2015} and experiment\cite{Zeiher2016,Labuhn2016}. The ultimate potential of Rydberg dressing remains unclear, as discussed in Sec. \ref{sec.dressing}, and more work is needed to fully understand the mechanism of and possible mitigation strategies for avalanche decay in a many body setting. 
Photonic quantum information processing mediated by Rydberg atoms has progressed rapidly\cite{Murray2016}, as reviewed elsewhere in this special issue\cite{Firstenberg2016}.
Also hybrid quantum interfaces with Rydberg interactions are being studied in many research labs. 

Not surprisingly there are many challenges to be solved before the dream of a neutral atom quantum computer becomes reality. Primary challenges include higher fidelity entangling gates, management of atom loading,  reloading of lost atoms, QND measurement of atomic states, low crosstalk qubit initialization and  measurement in arrays with few micron scale qubit spacings, and control of electric field noise near surfaces.

There is no principle reason that these challenges cannot be met. We have identified Rydberg gate protocols\cite{Theis2016} that can reach 
Bell state fidelity of $F=0.9999$ in real atoms at room temperature provided  
solutions to the challenges of Doppler dephasing, laser noise, and electric field noise described in Sec. \ref{sec.experimental} are met. 
Even higher fidelity should be possible in a cryogenic environment with increased Rydberg lifetime.
High fidelity control of a multi-qubit array and implementation of error correction will require closely integrated ultrahigh vacuum, electro-optical, laser, and classical computer hardware that is not available today and will need to be developed. The list may seem excessively daunting, but it is not more so than for other promising quantum computing technologies.

\ack

I am grateful for stimulating discussions with many colleagues, especially Dana Anderson, Ilya Beterov, Grant Biedermann, Antoine Browaeys, Jungsang Kim, Klaus M\o{}lmer,
Thad Walker, David Weiss, and Frank Wilhelm.  
This work was supported by the 
IARPA MQCO program through ARO 
contract  
W911NF-10-1-0347, the ARL-CDQI through cooperative agreement W911NF-15-2-0061, the AFOSR quantum memories MURI, and NSF  awards 1521374, 1404357.\\

%\section*{References}

\bibliographystyle{unsrt}
%%%\bibliographystyle{unsrtnat}
%%%\bibliographystyle{iopart-num}

%\bibliography{d:/users/mark/pubs/biblio/saffman-refs,d:/users/mark/pubs/biblio/rydberg_bib_v15,d:/users/mark/pubs/biblio/qc_refs%,d:/users/mark/pubs/biblio/atomic,d:/users/mark/pubs/biblio/optics,d:/users/mark/pubs/biblio/holmium_v1}

\end{document}